\documentclass[11pt,twoside]{article}

\usepackage{asp2006}
\usepackage{epsf}
\usepackage{psfig}
\usepackage{lscape}

\begin{document}
\title{Quasar Metal Abundances \& Host Galaxy Evolution}   
\author{Fred Hamann \& Craig Warner}   
\affil{University of Florida}    
\author{Matthias Dietrich}   
\affil{The Ohio State University}    
\author{Gary Ferland}
\affil{University of Kentucky}
\begin{abstract} 
Quasars signal a unique phase of galaxy evolution -- when massive spheroids 
are rapidly being assembled, forming stars and growing their central 
super-massive black holes. Measurements of the metal abundances around 
quasars provide unique information about these complex evolutionary 
processes. Here we provide a brief review of the current status 
and implications of quasar abundance research. 
\end{abstract}


\section{Goals \& Motivation} 
The central goal of quasar abundance studies is to understand the 
evolutionary relationship between 
quasars, super-massive black holes (SMBHs) and their host galaxies. 
We know that a close relationship exists because dormant 
SMBHs are not only common in galactic nuclei today, but their 
masses, $M_{BH}$, scale directly with the mass of the surrounding 
galactic spheroids, $M_{gal}$, (e.g., Tremaine et al. 2002). 
Whatever processes created the galactic spheroids must have also
(somehow) created central SMBHs with commensurate mass. Luminous quasars 
represent the final major growth stage(s) of the most massive SMBHs 
inside the most massive galaxies. The quasar phase is very brief, of order 
$10^8$ yr, but it coincides with a critical stage of SMBH--galaxy 
evolution when galactic spheroids are still rapidly being assembled and 
making stars. The energy output from accreting black holes might, 
in fact, regulate that star formation and thus lead  
naturally to the observed $M_{BH}$-- $M_{gal}$ correlation (Kauffmann \& 
Haehnelt 2000, Granato et al. 2004). 

Quasar abundance studies can help us understand better the complex 
evolutionary relationship between SMBHs and their host galaxies. 
For example, how much star formation (how much conversion of the initial 
gas into stars) occurs in galactic spheroids before the visible quasar 
epoch? Did the major star-forming episodes occur before, during or after 
the final luminous stages of SMBH growth? Is the relative 
timing of these events 
consistent with quasars triggering star formation, shutting it 
down, or having no affect at all? How much do outflows during the 
quasar epoch contribute to the distribution of metals to 
the surrounding galactic environments? Are the enrichment processes near 
quasars consistent with normal galactic chemical evolution, or do 
they require more exotic processes? Perhaps there are not single answers 
to these questions. A range of measured abundance results might imply 
a range of evolutionary/enrichment circumstances. On a more basic level, 
the gas-phase metal abundances can affect important aspects of quasar 
physics. For example, do higher metallicities (and thus higher 
opacities) lead to higher mass 
loss rates or gas ejection speeds from the central AGN? 

Reviews of quasar abundance work were given previously 
by Hamann et al. (2004) and Hamann \& Ferland (1999, hereafter HF99). 
Here we provide a very brief overview and an update on recent 
developments. 

\section{Metallicity Diagnostics \& Results}

We can infer the elemental abundances near quasars 
from a variety of spectroscopic 
observations. Each provides unique information. Each has limitations 
and uncertainties. It is important to consider all of them. 

\subsection{Broad Emission Lines}

One great advantage of the 
broad emission lines (BELs) for quasar abundance work is that 
they are relatively easy to measure in large numbers of objects. 
The first quantitative studies of BELs concluded that the 
metallicities are consistent with solar, with 
large ($\pm1$ dex) uncertainties (e.g., Shields 1976, 
Davidson \& Netzer 1979, Uomoto 1984). These studies had to deal 
with a problem that is pervasive in emission line analysis. The lines 
cool the gas where they form and, therefore, the total flux in all 
lines depends on the energy balance between heating and cooling and 
{\it not} (directly) on the metallicity. In practice, this means 
we cannot simply use the strength of a prominent metal line such 
as CIV $\lambda$1549 compared to a hydrogen line like Ly$\alpha$ 
to infer the C/H abundance ratio. However, as the metallicity goes 
up and the gas temperature goes down (which keeps 
the total line flux $\sim$constant), some of the relative line 
strengths do change. In addition, variations in the relative metal 
abundances can change specific line ratios as different 
lines carry a different share of the cooling (see Ferland et al. 
1996, Hamann et al. 2002). 

Shields (1976) noted that the relative abundance of nitrogen scales 
with metallicity in galactic stars and H~II regions as 
N/O $\propto$ O/H (also Pilyugin et al. 2003). 
Therefore, quasar BEL ratios like 
N~III] $\lambda$1750/O~III] $\lambda$1663 or N~III]/C~III] $\lambda$1909 
can be used to infer the metallicities because of their sensitivity 
to the N/O and N/C abundances. Shields (1976) chose to work with 
these relatively weak intercombination lines to avoid saturation and 
thermalization problems that might affect strong permitted lines. 
However, reverberation studies beginning in the 1980s showed that 
BEL region (BELR) densities are higher than previously believed 
(Ferland et al. 1992), and therefore the intercombination lines 
probably all form near their critical densities. Saturation and 
thermalization issues are unavoidable, but 
by the 1990s spectral simulations were up to the task. 

\begin{figure}[h]
\begin{center}
\vspace{-0.4cm}
\hbox{\hspace{0.1cm}
\psfig{figure=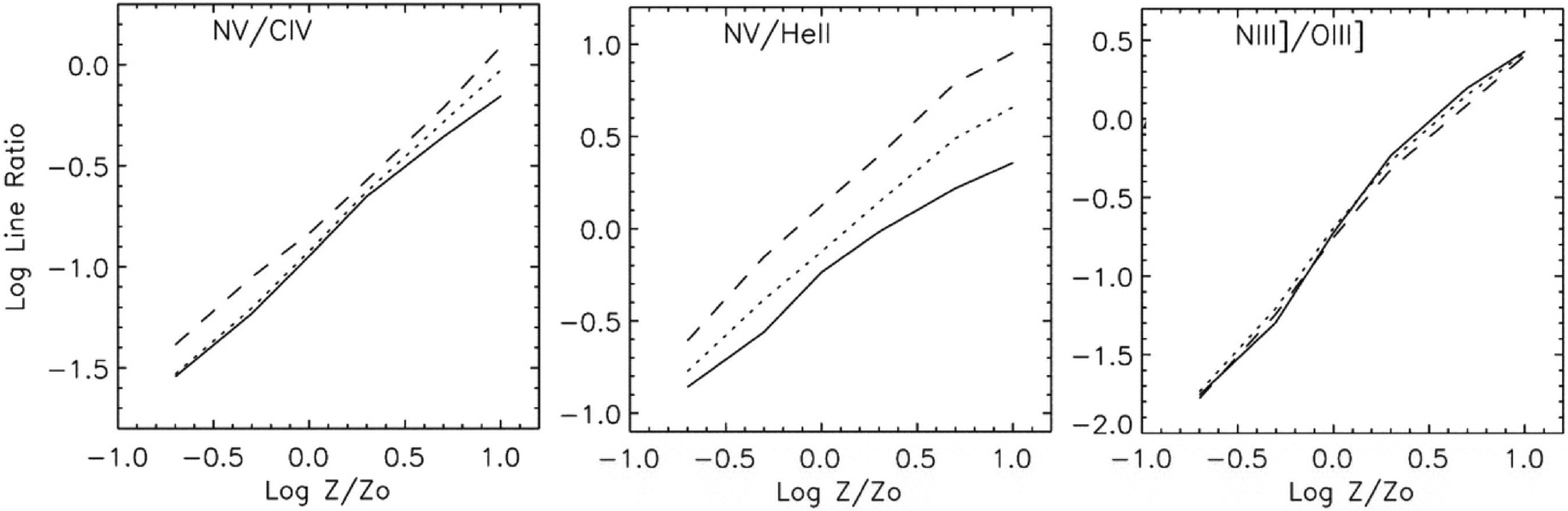,height=4.5cm,width=13.0cm,angle=0}
}
\vspace{-1.2cm}
\end{center}
\caption{\footnotesize
{Theoretical BEL ratios vs. metallicity for three different 
ionizing spectral shapes (solid, dashed, dotted lines) in the LOC model 
(Hamann et al. 2002).}}
\end{figure}

In a series of papers, Hamann \& Ferland (1992, 1993), Ferland 
et al. (1996) and HF99 performed 
extensive calculations with Gary Ferland's photoionization code 
CLOUDY to show that some of 
the stronger permitted line ratios, such as N V $\lambda$1240/C IV and 
N V/He II $\lambda$1640, are also sensitive to the metallicity if 
nitrogen behaves as a secondary element. Hamann et al. (2002) then 
extended that work to the Locally Optimally-emitting Cloud 
(LOC) model of the BELR (Baldwin et al. 1995). The LOC 
calculations are more realistic because they consider a stratified 
BELR, where clouds with a wide range 
of densities and incident fluxes all coexist. Each 
line forms preferentially wherever its emission is most favored, and 
the observed spectrum is the sum over many diverse BELR components. 
Hamann et al. (2002) used this model to calculate BEL strengths 
and ratios for a range of metallicities while scaling 
N/O $\propto$ O/H. Figure 1 shows some of their results. 
In one particular quasar at $z=4.16$, having many well-measured BELs, 
Warner et al. (2002) used this analysis to derive $Z_{gas}\sim 2$ 
Z$_{\odot}$. Figure 2 shows similar results for a sample of 70 
luminous high-redshift quasars (Dietrich et al. 2003). The average 
BEL metallicity inferred for that sample is $Z_{gas}\sim 4$ Z$_{\odot}$.
In the highest redshift quasar yet studied, at $z=6.28$, Pentericci et al. 
(2002) estimated $Z_{gas}\geq Z_{\odot}$ based on a measurement of 
N V/C IV and a lower limit on N V/He II. 
\begin{figure}[h]
\begin{center}
\vspace{0.0cm}
\hbox{\hspace{1.0cm}
\psfig{figure=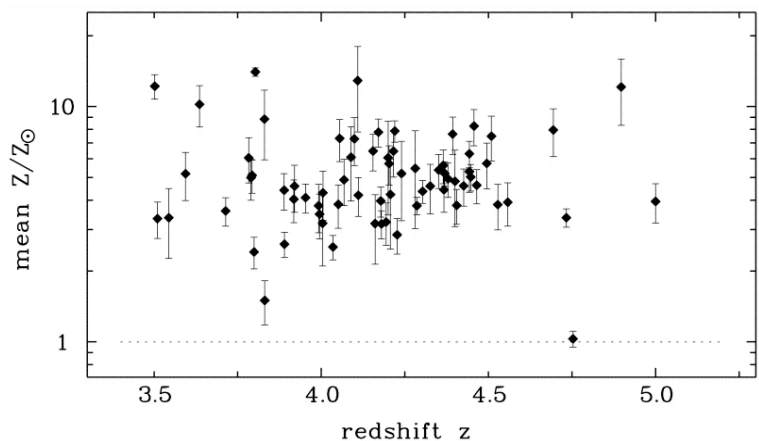,height=5.5cm,width=10.5cm,angle=0}
}
\vspace{-0.8cm}
\end{center}
\caption{\footnotesize
{Derived BELR metallicites, based on the average from several nitrogen 
line ratios, for 70 luminous high-redshift quasars (Dietrich et al. 2003).}}
\end{figure}

More recently, Nagao et al. (2006a) took the LOC abundance analysis one 
step farther. They included more UV line ratios and found the best overall 
fits to composite quasar spectra 
(sorted by redshift or luminosity) by varying both $Z_{gas}$ and the relative 
numbers of clouds with different physical parameters in the LOC calculations. 
The key advantages to this approach are i) less dependence on the assumed 
N/O $\propto$ O/H scaling, and ii) specific fits to each quasar/composite 
spectrum. It is reassuring that their analysis of $>$5000 SDSS quasar 
spectra yields results nearly identical to the previous studies (e.g., 
Figure 2). 

\subsection{Narrow Emission Lines}

Narrow emission lines (NELs) can provide a 
valuable test of the BEL results, although it is important to keep 
in mind that quasar NELs form much farther from the central engines,  
10$^2$ to 10$^4$ pc away compared to $<$1 pc for the BELs. 
The well-known NELs at rest-frame visible wavelengths have not 
yet been used to estimate abundances in high-redshift quasars because they 
are shifted into the near-IR and are thus more difficult to measure. 
However, there have been several studies of these lines in low-redshift 
AGN. For example, 
Groves et al. (2006) used photoionization models and standard line 
ratios such as [O III] $\lambda$5007/H$\beta$ and [N II] 
$\lambda$6583/H$\alpha$ to infer NEL region metallicities for $\sim$23,000 
Seyfert 2s in the SDSS database. They adopt a single set of physical 
parameters for the entire sample and, remarkably, 
find that all but $\sim$40 of the $\sim$23,000 sources (99.8\%) 
have $Z_{gas}\geq Z_{\odot}$. The typical metallicity is roughly 
2--4 $Z_{\odot}$. These results are consistent with eariler NEL 
studies that applied similar analysis 
to much smaller samples (Storchi Bergmann \& Pastoriza 1989, Storchi 
Bergmann et al. 1998, Groves et al. 2004). 

Nagao et al. (2006b) took a different approach to the NEL analysis. 
They examined high-redshift AGN that have only narrow lines (e.g., 
QSO 2s and narrow-line radio galaxies) so they could measure the same 
rest-frame UV lines used in the BEL analysis. They found a 
wider range of NEL metallicities than Groves et al. (2006), 
from 0.2 to 5 $Z_{\odot}$, and they note that the results depend on the 
assumed the gas density. 
Both Groves et al. (2006) and Nagao et al. (2006b) performed 
1-zone spectral synthesis calculations and used the same N/O $\propto$ O/H 
scaling as Hamann et al. (2002). 

\subsection{Associated Absorption Lines}

Quasar absorption lines provide another opportunity 
to measure the element abundances. 
Most of the work in this area has been on the 
narrow ``associated" absorption lines (AALs, near the emission 
redshifts) because i) most of them are physically related to the 
quasar or its host galaxy (e.g., Nestor et al., this proceedings), 
and ii) the line profiles are narrow enough to resolve important 
doublets, such as C IV $\lambda$1548,1551, and thereby diagnose 
potential saturation problems. In principle, the absorption line 
analysis is much simpler 
than the emission lines because one needs only to measure the ionic 
column densities and apply ionization corrections to derive 
abundance ratios. No assumptions about secondary N are needed.
In practice, however, 
the ionization corrections can be uncertain and the absorber 
geometries can be complex (e.g., with partial coverage of the 
background emission source, see Hamann 1997 and Hamann et al. 2004, 
Gabel et al. 2006). 
One also has to take care in interpreting the AAL results because 
the lines can form in a wide range of locations, often poorly 
known but ranging from winds very near the quasars to extended 
halo gas $>$100 kpc distant (Simon et al. 2007, 
this proceedings). 

Progress with the absorption lines has been slow because 
it requires high resolution spectra obtained with  
large telescopes. The first reliable studies in the 1990s 
found typically $Z_{gas}\geq Z_{\odot}$ with N/C also enhanced, 
consistent with the emission line data  
(see HF99 and refs. therein). More recent studies find similar results. 
For example, D'Odorica et al. (2004) carefully selected quasars measured 
with the UVES echelle spectrograph at the VLT ($\sim$7 km/s resolution)  
to estimate AAL abundances in six well-measured quasars at redshifts between 
2.1 and 2.6. Five of these six have $Z_{gas}$ = 1 to 3 Z$_{\odot}$. 
The current state of the art in this analysis is illustrated by 
Gabel et al. (2006), who estimated $Z_{gas}$ to be several 
times solar with nitrogen 2 to 3 times more enhanced in a 
quasar at redshift $\sim$2.2 (see also Arav et al. 2007). 
The absorption line data thus continue to 
support the emission line results. 
In one study that compares directly the AAL and BEL data, 
Kuraszkiewicz \& Green (2002) showed the ratios of N~V/C~IV line 
strengths in 
emission and absorption correlate with each other in a sample 
of 17 low-redshift quasars. 

\section{How Much Metal-Rich Gas Is There?}

There are two important questions not addressed directly by the 
quasar metallicity data: How much metal-rich gas is present near 
quasars, and what stellar populations enriched this gas? We could 
start by looking at the mass of gas in the BELRs, $M_{BELR}$. 
LOC models indicate that luminous quasars have typically 
$M_{BELR} \sim 1000$ M$_{\odot}$ (Baldwin et al. 2003). However, 
this is an extreme lower limit to the total mass of metal-rich 
gas, $M_{gas}$, because BELRs are dynamic regions that 
are probably continuously replenished by material discharged 
from the accretion disk. The total reservoir of gas sampled 
by the BELR during a quasar's lifetime is therefore 
the total mass of matter accreted through the disk, roughly $M_{BH}\sim 
10^9$ M$_{\odot}$. This is still just a lower limit to $M_{gas}$, however,  
because not all of the metal-rich gas around quasars 
will funnel through the accretion disk. Therefore the mass in metal-rich 
gas is $M_{gas}\geq M_{BH}$. The stellar populations needed to enrich 
this gas to super-solar levels should have even larger masses, 
nominally $M_{stars}\sim$ few $\times 
\, M_{gas}$ (for standard stellar yields and initial mass functions). 
We conclude that stellar populations with mass 
$M_{stars} >$ few $\times 10^9$ M$_{\odot}$ are already present or 
perhaps still rapidly forming when luminous quasars become 
optically observable. 

\section{Other Enrichment/Star Formation Indicators}

Molecular line and far-IR through mm continuum observations can also 
probe the gas-phase metal abundances and 
stellar populations near quasars. For example, Cox et al. (2006) 
estimate that $\sim$30\% of high-redshift optically luminous quasars 
qualify as Ultra-Luminous Infrared Galaxies (ULIRGs) based on their 
rest-frame far-IR and sub-mm luminosities (also Beelen et al. 2006). 
These sources have enourmous star formation rates of order $\sim$1000 
M$_{\odot}$/yr, assuming the IR luminosity is not powered by the quasars. 
(There is evidence that this is so when the observed far-IR/radio 
flux ratios are similar to nearby starbursting galaxies without AGN.) 

These are important constraints on the rates of star formation that 
can be coincident with the brief visible quasar phase. But another 
important constraint is often overlooked. The observed masses of CO 
or FIR-emitting dust directly constrain the amount 
of star formation that {\it preceded} the quasar epoch. For example,
the dust masses inferred from the studies 
mentioned above are typically $10^8$ to $10^9$ M$_{\odot}$, which 
corresponds to total masses in metal-rich gas of $M_{gas}\approx 
10^{10}$($Z_{gas}$/Z$_{\odot}$) to $10^{11}$($Z_{gas}$/Z$_{\odot}$)  
M$_{\odot}$, where $Z_{gas}$/Z$_{\odot}$ is the normalized metallicity. 
To create these huge amounts of metals, there must 
be correspondingly massive stellar populations already in 
place {\it before} the observed quasar epochs. 
With nominal stellar yields and 
initial mass functions, the mass in stars already 
formed around these high-redshift quasars should be roughly 
$M_{stars} \sim$ few $\times\, 10^{10}$ to $10^{12}$ M$_{\odot}$. 

\section{Trends with $L$, $M_{BH}$, $L/L_{Edd}$ or Host Galaxy Properties}

One important result from quasar metallicity work is that there is 
no significant trend with redshift. 
This can be seen, for example, in Figure 2 and in other 
BEL studies that span a wider range in $z$ (HF99, Dietrich et al. 2003, 
Nagao et al. 2006a). There are far fewer results available from quasar 
absorption lines, but they too show no apparent trend with redshift. 
BEL studies do, however, indicate that there is a strong trend 
(apparently at every redshift) for larger nitrogen line ratios 
(higher metallicities) in more luminous quasars 
(HF99, Warner et al. 2003, Nagao et al. 2006a). 

What is the physical basis for this $L$--$Z$ 
relationship? Shemmer et al. (2004) examined the N~V/C~IV BEL ratios 
in a sample of 92 AGN which had H$\beta$ available to measure 
SMBH masses. They found that the N~V/C~IV line ratio correlates 
more strongly with the normalized accretion rate, $L/L_{Edd}$, 
than it does with $L$ or $M_{BH}$. The sense of the relationship is 
for higher metallicities at larger $L/L_{Edd}$. Shemmer et al. 
speculate that this trend might be indicative of higher $Z$ 
at earlier evolutionary stages, when perhaps $L$, $L/L_{Edd}$ and the 
star formation rates are all larger. 

In a similar study, Warner et al. (2003, 2004, 2007) examined the nitrogen 
BEL ratios and used C~IV to derive M$_{BH}$ in a sample of 578 AGN. 
They do not find a significant relationship to $L/L_{Edd}$, but 
instead find a strong trend for increasing metallicity with SMBH mass. 
(This departure from the Shemmer et al. 2004 is not understood, 
but it might derive from differences in their quasar samples.)
To isolate the dependences on $L$ and $M_{BH}$, Warner et al. 
(2007) examined quasar sub-samples that span i) 
a range in $L$ at nearly constant $M_{BH}$, and ii) a 
range in $M_{BH}$ at nearly constant $L$. The main results are shown 
in Figure 3. There is a strong trend for increasing nitrogen BEL ratios 
(metallicity) with larger $M_{BH}$ even when $L$ is held fixed. But there 
is no trend at all with $L$ if $M_{BH}$ is constant. Warner et al. 
attribute this underlying 
$M_{BH}$--$Z$ correlation to a mass--metallicity trend 
in quasar host galaxies. More massive SMBHs reside in more massive 
spheroids, which characteristically have higher metallicities (Trager 
et al. 2000a). In fact, 
Warner et al. (2003) showed that the slope of the $M_{BH}$--$Z$ 
relationship in their quasar sample roughly mimics the slope in 
the galactic mass--metallicity relation. 
Further evidence for a mass--metallicity trend in AGN comes from the 
Groves et al. (2006) study of low-redshift Seyfert 2 galaxies (\S2.2). 
They inferred host galaxy masses directly from the SDSS imaging data and 
found that the more massive galaxies contain systematically 
more metal-rich NEL regions. 
\begin{figure}[h]
\begin{center}
\vspace{-0.5cm}
\hbox{\hspace{-0.1cm}
\psfig{figure=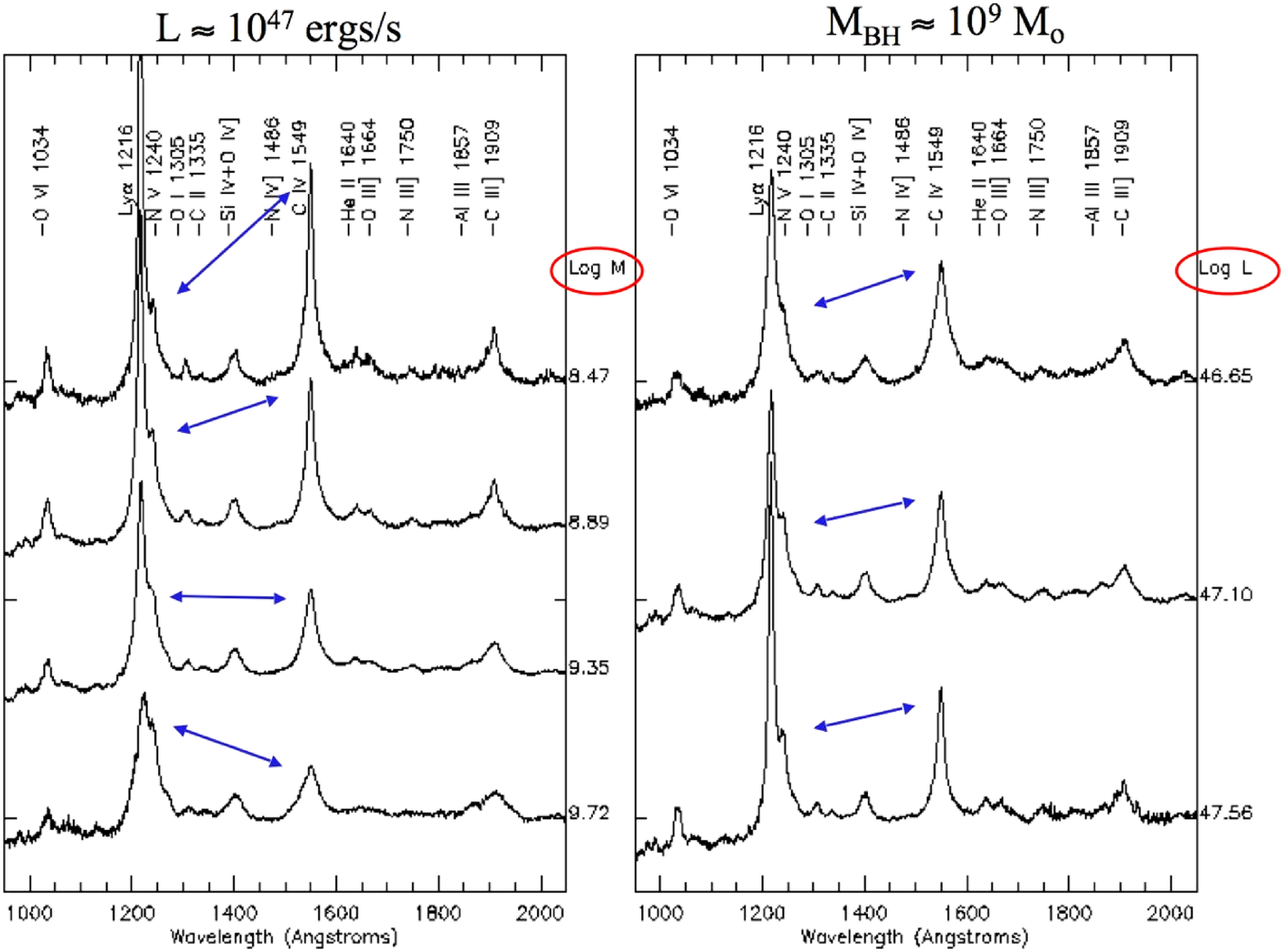,height=10.5cm,width=13.5cm,angle=0}
}
\vspace{-1.5cm}
\end{center}
\caption{\footnotesize
{Composite spectra of quasars spanning a range in SMBH mass  
at a fixed luminosity ($L\approx 10^{47}$ ergs/s, left-hand plot), 
and a range in luminosity at fixed SMBH mass ($M_{BH}\approx 10^9$ 
M$_{\odot}$, right-hand plot). Each 
composite includes at least 30 spectra, with the 
average values of $L$ and $M_{BH}$ shown at the right. 
The spectra are normalized and shifted vertically for easier display. 
There is a strong trend in the nitrogen line ratios, 
including NV/CIV marked 
by arrows, for increasing metallicity with increasing 
$M_{BH}$, but no trend at all with $L$. From Warner et al. (2007).
}}
\end{figure}

\section{Implications for SMBH--Host Galaxy Evolution}

What do the results for $Z_{gas}\geq$ Z$_{\odot}$ near quasars tell us 
about the coupled evolution of galaxies and SMBHs? 
The first thing to realize is that the quasar metallicities 
are consistent with normal galactic chemical evolution. 
In particular, the stars in massive spheroids today are mostly old and 
metal rich. They formed typically 2--12 Gyr ago and attained mean 
metallicities near their cores of $\left<Z_{stars}\right> = 
1$--3 Z$_{\odot}$ (e.g., Figure 4). 
These are the environments that today contain high-mass SMBHs and were 
once active quasar hosts. Most of the star formation occurred 
abruptly, within 0.5 Gyr, based on sub-solar ratios of 
Fe/$\alpha$-element abundances in the present-day stars. 
Near the end of the main star-forming epoch, the gas that created these  
stars {\it must} have had 
metallicities $Z_{gas} > \left<Z_{stars}\right>$ and quite possibly 
2 to 3 times higher. Figure 4 shows a simple chemical evolution 
model that illustrates the essential features (Friaca \& Terlevich 1998). 
In this typical simulation, the star formation is nearly complete 
(with $\geq$70\% of the initial converted into stars) 
after just $\sim$0.5 Gyr when $Z_{gas}\approx 4$ Z$_{\odot}$. 
Nitrogen is selectively enhanced by secondary enrichment. 

\begin{figure}[h]
\begin{center}
\vspace{-0.2cm}
\hbox{\hspace{-0.3cm}
\psfig{figure=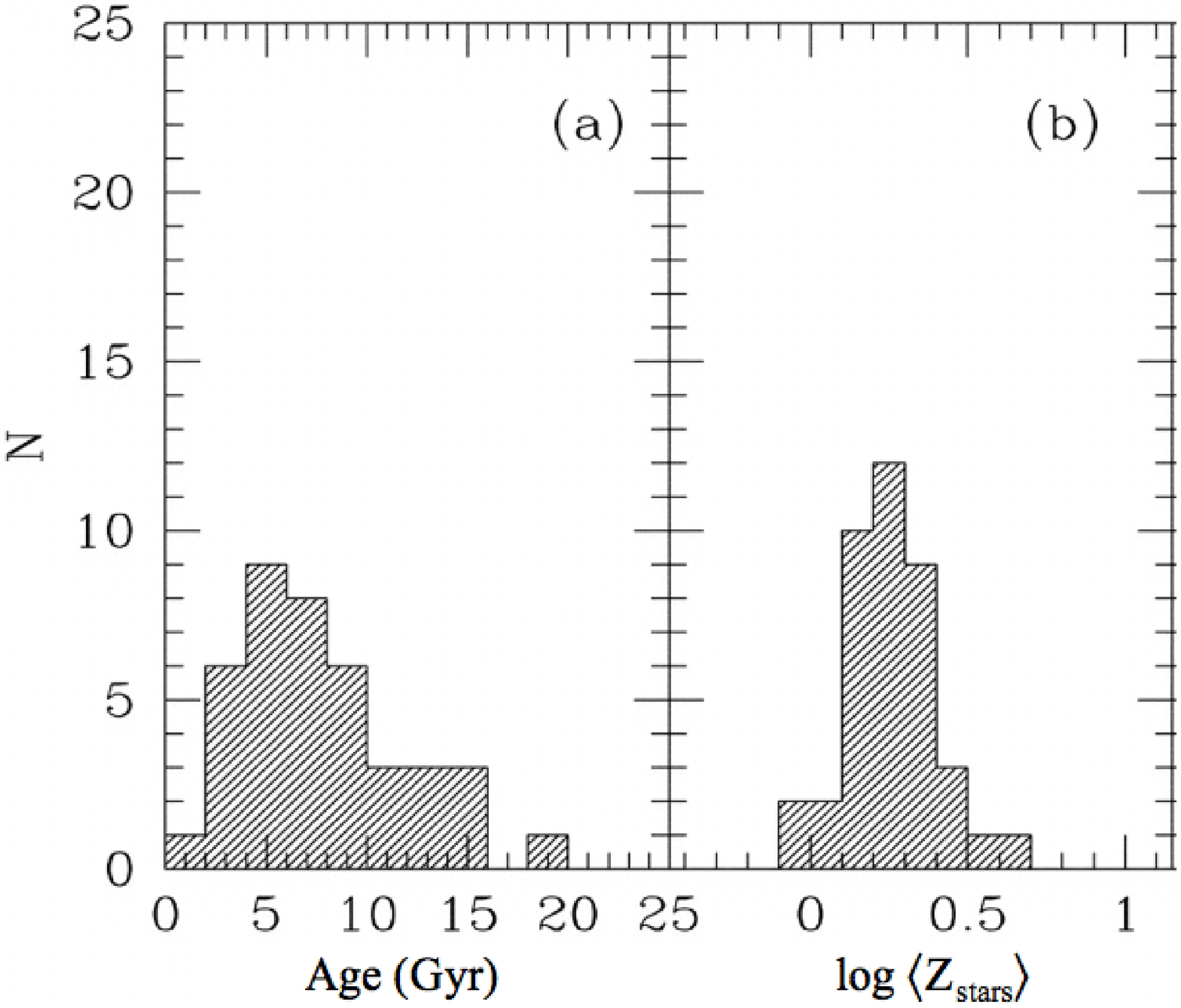,height=5.4cm,width=6.7cm,angle=0}
\psfig{figure=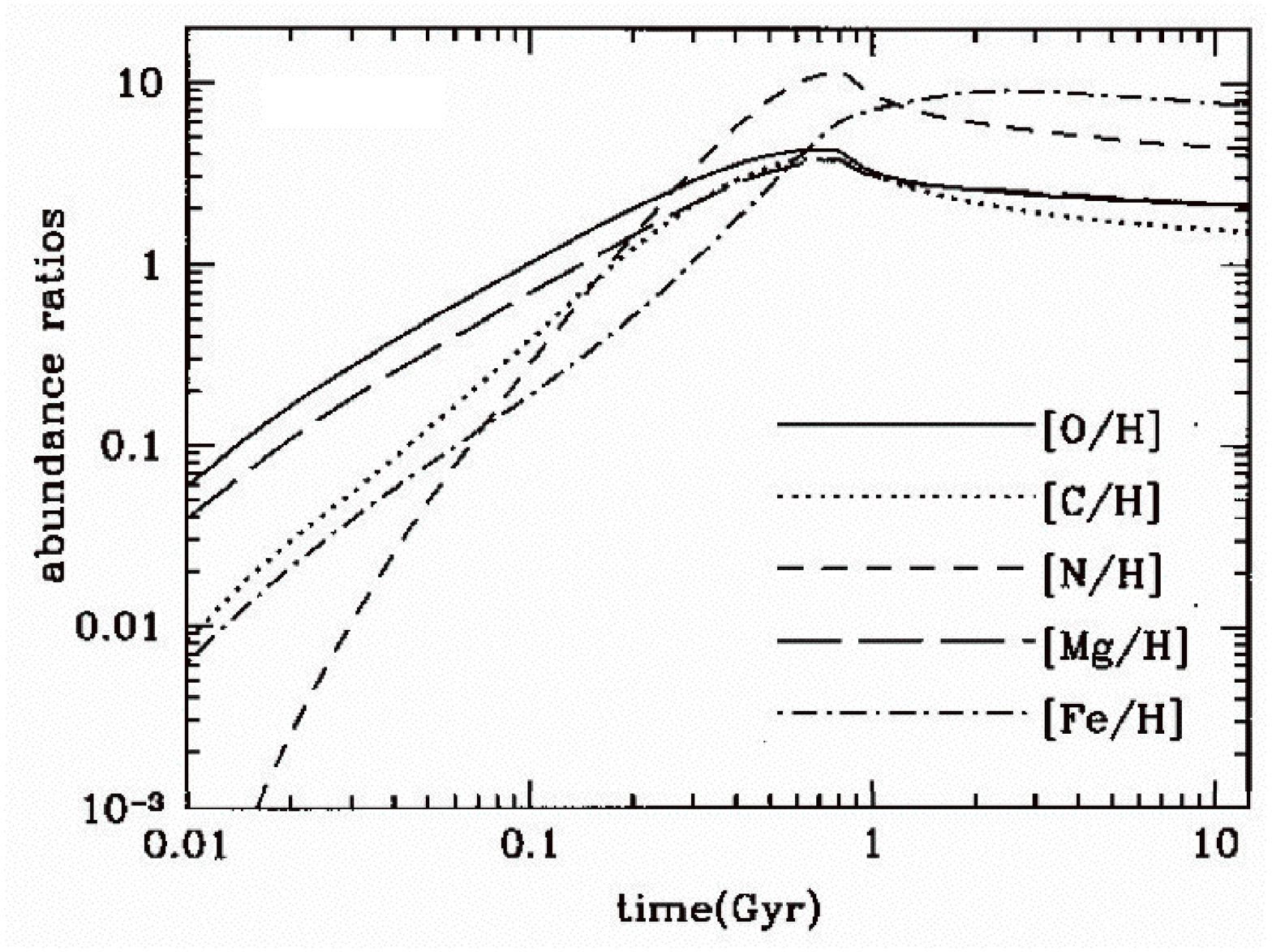,height=5.4cm,width=6.7cm,angle=0}
}
\vspace{-1.2cm}
\end{center}
\caption{\footnotesize
{{\it Left-hand panels:} Observed mean stellar ages and metallicities in the 
central regions ($r_e/8$) of field elliptical galaxies (adapted from Trager et 
al. 2000b). {\it Right:} Abundance ratios normalized to solar in a typical 
1-zone + infall chemical evolution model of ellipticals (from Friaca \& 
Terlevich 1998). [O/H] best represents the total metallicity. Nitrogen is 
selectively enhanced by secondary enrichment. The rise in [Fe/H] is 
delayed by the contribution from SN Ia's.
}}
\label{fig1}
\end{figure}
Quasar abundance data provide important new information. 
In particular, the preponderance of metal-rich quasars, and the large 
masses of dust or CO found near some of them, mean that the 
star-formation in galactic spheroids was largely complete, 
or at least well underway, when the central quasars became visibly 
luminous objects. The quasar data also support the idea that more 
massive galaxies have more extensive star formation (converting more 
of their gas into stars), leading to higher metallicities, e.g., 
in their central regions. 

These quasar results agree well with recent physically-motivated 
models of galaxy--SMBH evolution that try, in particular, to explain 
the observed $M_{BH}\propto M_{gal}$ relationship (Kauffmann \& 
Haehnelt 2000, Granato et al. 2004, DiMatteo 
et al. 2004, Hopkins et al. 2005, Springel et al. 2005). 
In those models, major mergers of gas-rich proto-galaxies trigger 
enormous bursts of star formation and funnel gas towards a central 
SMBH. Feedback from the growing SMBH, and to a lesser extent supernovae 
in the starburst, eventually halts the star formation and produces the  
$M_{BH}\propto M_{gal}$ correlation. The central SMBH is under-luminous 
and obscured by dust during the early stages. It appears to us 
as a visibly luminous quasar only {\it after} the major starburst phase. 
The models predict gas-phase metallicities near the quasars 
of typically 2--3 Z$_{\odot}$. Figure 5 shows the results of 
one particular simulation compared to the highest redshift quasar known 
to date, at $z\approx 6.5$ (Li et al. 2006). 
In this extreme case, multiple mergers lead to a 
total star formation rate that briefly exceeds $10^4$ M$_{\odot}$/yr at 
$z\sim 8.7$ and creates a total stellar mass of $\sim$$10^{12}$ M$_{\odot}$ 
{\it before} the quasar begins to dominate the luminosity and emerge as a 
visibly bright object at $z\sim 6.5$. The final black hole mass is 
$M_{BH}\sim 2\times 10^9$ M$_{\odot}$. 

\begin{figure}[h]
\begin{center}
\vspace{-0.2cm}
\hbox{\hspace{2.0cm}
\psfig{figure=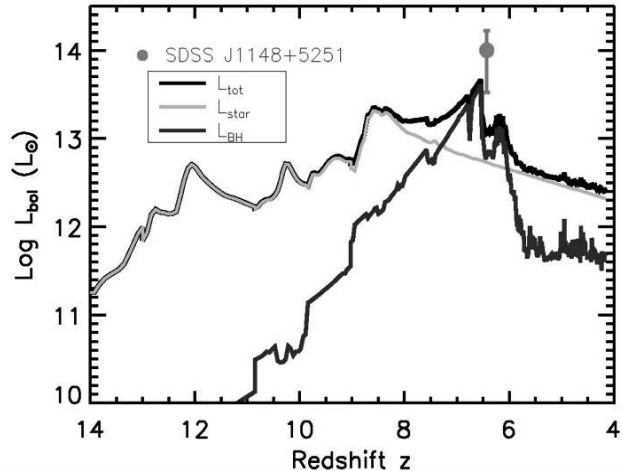,height=6.5cm,width=8.5cm,angle=0}
}
\vspace{-1.2cm}
\end{center}
\caption{\footnotesize
{Evolution in the total, stellar and SMBH bolometric luminosities 
in a galaxy/SMBH formation model compared to the highest redshift 
quasar known, SDSS J1148+5251 (bold dot with error bar, from Li et 
al. 2006). A bright, blue, unobscured quasar appears in the simulation 
around redshift 6.5, roughly 
0.3 Gyr after the star formation peak at redshift $\sim$8.7.
}}
\end{figure}

\section{Summary \& Future Prospects}
The main results are the following. 1) Quasar environments are 
metal-rich, with typically $Z_{gas}\sim 1$--5 Z$_{\odot}$, 
at all observed redshifts. 2) They 
were enriched by stellar populations with at least Galactic bulge-like 
stellar 
masses. 3) The major star-forming episodes in galactic spheroids must have 
occurred before the observed quasar epochs, probably with most of the  
stellar mass involved in the $M_{BH}$--$M_{gal}$ relation already in 
place when quasars first appear as a visibly luminous objects. 4) The 
ensemble results support an 
evolutionary sequence like this: major merger $\rightarrow$ ULIRG/starburst 
$\rightarrow$ possible transition object (declining star formation rate or 
partially obscured AGN) $\rightarrow$ visible quasar. They are also 
consistent with the idea that SMBH activity halts the star formation 
and thus leads to the observed 
$M_{BH}$--$M_{gal}$ relation. 5) Quasars in more massive hosts 
are more metal-rich, consistent with a normal galactic mass--metallicity 
relation with perhaps an additional dependence on $L/L_{Edd}$ related 
to the quasar's age.

More work is clearly neeeded. For 
example, 1) we need more measurements of AAL and NEL abundances, e.g., 
at high redshifts, to test the current BEL results. 2) We should 
compare quasar abundance data directly to 
host galaxy properties, such as mass, age and metallicity of the 
stellar populations, to understand better the evolutionary 
sequence of merger-to-quasar outlined above. 3) Transition 
objects (with stronger FIR/sub-mm emission or perhaps redder quasar 
spectra) deserve more attention because they might represent a unique 
evolutionary phase with younger or more obscured AGN inside younger 
galaxies. 4) Larger data samples are needed to probe further the relationships 
of quasar abundances and associated star formation with mass and 
$L/L_{Edd}$. Finally, 5) measurements of specific element ratios 
in quasars, most notably Fe/$\alpha$, will provide more information 
on the nature and timescales of the star formation and chemical 
enrichment. 

\acknowledgements 
This work was supported by grants from the National Science Foundation 
(AST99-84040) and STScI (GO-09871.02).
\bigskip\smallskip

\noindent{\bf References}
\medskip



\baselineskip 13pt
\parskip=0pt
\leftskip=0.5in
\parindent=-0.5in

Arav, N., et al. 2007, ApJ, in press

Baldwin, J.A., et al.
2003, \apj, 582, 590

Baldwin J., Ferland G., Korista K.T., \& Verner D. 1995, \apj, 455,
L119


Beelen, A., et al. 2006, ApJ, 642, 694


Cox, P., et al. 2006, in ``The Dusty and Molecular Universe," ESA SP-577, 115

Davidson, K., \& Netzer H. 1979, Rev. Mod. Physics, 51, 715

Dietrich, M., Hamann, F., et al. 2003, 589, 722

DiMatteo, T., et al. 2004, ApJ, 610, 80

D'Odorico, V., et al. 2004, MNRAS, 351, 976

Ferland, G. J., et al. 1992, ApJ, 387, 95

Ferland, G. J., et al. 
1996, ApJ, 461, 683


Fria\c{c}a, A.C.S. \& Terlevich, R.J. 1998, \mnras, 298, 399

Gabel, J. R., Arav, N., \& Kim, T.-S. 2006, ApJ, 646, 742


Granato, G. L., et al.
2004, ApJ, 600, 580

Groves, B., Heckman, T. M., \& Kauffmann, G. 2006, MNRAS, 371, 1559

Hamann, F. 1997, ApJS, 109, 279

Hamann, F., \& Ferland, G. J. 1992, ApJL, 391, L53

Hamann, F., \& Ferland, G. J., 1993, ApJ, 418, 11


Hamann F. \& Ferland G. 1999, ARAA, 37, 487

Hamann, F., et al.
2002, ApJ, 550, 142.

Hamann, F., et al. 
2004, Carnegie Observatories Astrophysics Series, 4, 440

Hopkins, P. F., et al. 2005, ApJ, 630, 705

Kauffmann, G., \& Haehnelt, M. 2000, MNRAS, 311, 576

Kuraszkiewicz, J. K., \& Green, P. J. 2002, ApJ, 581, 77

Li, Y., et al. 2006, ApJ, in press (astro-ph/0608190)



Nagao, T., Marconi, A., Maiolino, R. 2006a, AA, 447, 157

Nagao, T., Marconi, A., Maiolino, R. 2006b, AA, 447, 863



Pentericci, L., et al. 2002, AJ, 123, 2151

Pilyugin, L.S., Thuan, T.X., Vilchez, J.M. 2003, AA, 397, 487


Schemmer, O., et al. 2004, ApJ, 614, 547

Shields, G. A., 1976, ApJ, 204, 330

Springel, V., DiMatteo, T., \& Hernquist, L. 2005, MNRAS, 361, 776

Storchi-Bergmann, T., \& Pastoriza, M. G. 1989, ApJ, 347, 195

Storchi-Bergmann, T., et al. 1998, AJ, 115, 909

Trager, S., et al. 2000b, AJ, 119, 1645

Trager, S., et al. 2000a, AJ, 120, 165

Tremaine, S., et al. 2002, \apj, 574, 740

Uomoto, A. 1984, ApJ, 284, 497


Warner, C., et al. 2002, ApJ, 567, 68

Warner, C., Hamann, F., \& Dietrich, M. 2003, ApJ, 596, 72

Warner, C., Hamann, F., \& Dietrich, M. 2004, ApJ, 608, 136

Warner, C., Hamann, F., \& Dietrich, M. 2007, ApJ, submitted


\end{document}